# Formation of Massive Counterrotating Disks in Spiral Galaxies

Aniruddha R. Thakar and Barbara S. Ryden[1]

Department of Astronomy, The Ohio State University, Columbus, OH 43210-1106

email: thakar.4@osu.edu; ryden@payne.mps.ohio-state.edu

http://www-astronomy.mps.ohio-state.edu/

## ABSTRACT

We present results of numerical simulations of the formation of a massive counterrotating gas disk in a spiral galaxy. Using a hierarchical tree gravity solver combined with a sticky-particle gas dissipation scheme for our simulations, we have investigated three mechanisms: episodic and continuous gas infall, and a merger with a gas-rich dwarf galaxy. We find that both episodic and continuous gas infall work reasonably well and are able to produce a substantial gas counterrotating disk without upsetting the stability of the existing disk drastically, but it is very important for the gas to be well-dispersed in phase-space and not form concentrated clumps prior to its absorption by the disk galaxy. The initial angular momentum of the gas also plays a crucial role in determining the scale length of the counterrotating disk formed and the time it takes to form. The rate of infall, i.e. the mass of gas falling in per unit time, has to be small enough to preclude excessive heating of the preexisting disk. It is much easier in general to produce a smaller counterrotating disk than it is to produce an extensive disk whose scale length is similar to that of the original prograde disk.

A gas-rich dwarf merger does not appear to be a viable mechanism to produce a massive counterrotating disk, because only a very small dwarf galaxy can produce a counterrotating disk without increasing the thickness of the existing disk by an order of magnitude, and the time-scale for this process is prohibitively long because it makes it very unlikely that several such mergers can accumulate a massive counterrotating disk over a Hubble time.

*Subject headings:* galaxies: spiral — galaxies: structure — galaxies: evolution — galaxies: interactions — galaxies: kinematics and dynamics — hydrodynamics

---

## 1. Introduction

In recent years, high spectral resolution combined with sophisticated computer spectral analysis techniques has yielded line-of-sight velocity distributions of galaxies which reveal the vast kinematic complexity shown by the stars and gas in some elliptical, lenticular and spiral galaxies. One of the most interesting phenomena is the presence of counterrotating disks, i.e. disks rotating in a sense opposite to that of the galaxy as a whole. In ellipticals, the disks are usually confined to the nuclear regions, forming counterrotating *cores*. In some S0/spiral galaxies, however, up to half of the stars and/or gas in the disk system can be in retrograde orbits, forming counterrotating *disks* that extend far beyond the nuclear regions, e.g. NGC 4550 (Rubin *et al.* 1992), NGC 4826 (Braun *et al.* 1992), NGC 4546 (Sage & Galletta 1994), NGC 7217 (Merrifield & Kuijken 1994), and most recently NGC 3626 (Ciri *et al.* 1995).

The counterrotating cores seen in ellipticals seem to be the result of dissipationless, retrograde mergers between progenitors of different sizes (Kormendy 1984). Simulations of mergers have shown that the core of the secondary galaxy sinks to the center of the primary as a result of dynamical friction in such a merger (Balcells & Quinn 1990, Barnes & Hernquist 1991, Hernquist & Barnes 1991). Counterrotating disks in spiral galaxies, however, are intriguing because both disks can be cold and have similar scale lengths, as in NGC 4550 (Rix *et al.* 1992). It is very hard to imagine such disks being formed together, although it has been suggested that two *identical* counterrotating disks could conceivably arise due to the conversion of box orbits to tube orbits when a triaxial halo potential becomes axisymmetric (Evans & Collett 1994). Box orbits are equally likely to scatter in either direction, giving rise to equal populations of stars orbiting prograde and retrograde. In the general case, however, where the counterrotating disk is of a different size, has a significantly different velocity dispersion, or is inclined to the plane of the primary disk, one of the disks is almost certainly of external and secondary origin.

However, a merger between two disk galaxies with comparable masses but opposite spins would not be able to produce a counterrotating disk. Simulations of such mergers (e.g. Barnes 1992) have established that the tidal trauma associated with them plays havoc with the velocity fields of both disks, and the final product looks very much like an elliptical galaxy. It is much more likely that the counterrotating disk is formed by secondary infall of gas which was initially moving in a retrograde orbit and which was forced to settle in a flat disk aligned with the potential symmetry plane due to energy dissipation and subsequent loss of angular momentum (Rix *et al.* 1992). The preexisting disk can remain unperturbed if it has little or no prograde gas to begin with, and if the gas build-up is adiabatic. This is especially relevant in the light of cosmological models which predict that galaxies continue to accrete matter throughout their lives (Ryden & Gunn 1987, Ryden 1988, White 1990), and that the spin angular momentum of a galaxy with respect to this infalling material changes on time-scales much smaller than the galaxy's lifetime (Ryden 1988, Ostriker & Binney 1989). Spirals must also have an ongoing supply of outside gas if they are to maintain their observed star-formation rates over times comparable to the age of the universe (White 1991). Thus it is quite possible that, over time, infalling gas may assume a retrograde orbit around a galaxy and settle gradually into a counterrotating disk as large as the existing disk.

Although dissipationless mergers may be safely ruled out by the observed properties of counterrotating disks in spirals, it is not so easy to dismiss retrograde mergers with gas-rich dwarf galaxies in which dissipation does play a significant role. The ubiquity of dwarf galaxies (especially in clusters) suggests that this may be a potentially frequent phenomenon and worth investigating. Dwarf galaxies generally contain a fraction (typically a tenth) of the mass of normal galaxies. In a gas-rich dwarf, most of the luminous mass is in the form of gas (Roberts & Haynes 1994). The stellar mass is thus only a few percent of the mass of the original galaxy, and therefore likely to be much less disruptive to the spiral galaxy's disk. Dwarf galaxies also have dark halos, and the question is whether the stellar and gas components of the dwarf can form a counterrotating disk in spite of the dark halo and its potentially disruptive influence on the primary disk.

We have undertaken the task of developing hydrodynamical N-body simulations to investigate the dissipational infall and dwarf-merger hypotheses for forming counterrotating disks. Our numerical code consists of two components: a hierarchical tree gravity solver and a simple gas dissipation scheme. The gravity solver is a specialization of the generic TREE code (Barnes & Hut 1986) which has emerged as the method of choice among N-body codes in the last few years due to the gridless flexibility it affords and the excellent blend of speed and accuracy it provides for most applications. The preferred method of incorporating gas dynamics into N-body simulations currently is Smoothed Particle Hydrodynamics (Lucy 1977, Gingold & Monaghan 1977, Monaghan 1992), which allows realistic treatment of the physics involved and also treats the gas as a continuous medium rather than discrete particles. However, crude gas dynamics which represent gas clouds as particles and dissipate energy via inelastic collisions between these clouds can be used as a short-cut to gain some quick insights. Such "sticky particle" schemes have proven to be quite useful in the past (Brahic 1977, Schwarz 1981, Negroponte & White 1983, Roberts & Hausman 1984, Shlosman & Noguchi 1993), and in fact they may represent the clumpy nature of the gas better than SPH. We have therefore adopted this approach as a first approximation in order to test our galaxy model and study the macroscopic behavior of the gas prior to implementing "real" gas dynamics with SPH.

Our physical model and numerical method are described in detail in the next section, followed by a discussion of the simulation results and their implications in section 3. Section 4 summarizes our conclusions.

## 2. Computational Method

### 2.1. Spiral Galaxy Model

We adopt a two-component spiral galaxy consisting of a disk and a halo. The disk mass distribution is exponential in both the radial and vertical directions. The initial velocities are assigned using the epicycle approximation. The shape and orientation of the Schwarzschild velocity ellipsoid in the plane of the disk is assumed to be constant so that the radial, azimuthal and vertical dispersions bear a constant ratio to each other everywhere in the plane. The radial dispersion $\sigma_r$ is assumed to be proportional to the local disk surface density and is set by specifying the value of the stability parameter $Q$. The azimuthal dispersion $\sigma_\phi$ is obtained from $\sigma_r$ and standard epicycle theory (Binney & Tremaine 1987). For an exponential vertical mass distribution, the vertical dispersion $\sigma_z(z)$ can be obtained by integrating the Boltzmann equation for the plane-parallel approximation (van der Kruit 1988):

$$\sigma_z^2(z) = \sigma_z^2(0)\{2 - \exp(-z/z_e)\},$$
$$\sigma_z^2(0) = \pi G \Sigma z_e$$

where $\Sigma$ is the disk surface density and $z_e$ is the disk scale height. The rotation velocity is then calculated according to standard procedure from the circular speed and the velocity dispersions by solving Jeans equation (Binney & Tremaine 1987).

The halo mass distribution is that of a truncated isothermal sphere supported by a Maxwellian random velocity distribution (truncated at $2\sigma$ to avoid excessive evaporation). The halo is assumed to be four times as massive as the disk. The disk and halo are dissipationless and assumed to contain no gas initially.

Our units are $[t] = 1$ Myr, $[m] = 2.2 \times 10^{11}$ M$_\odot$, and $[r] = 1$ kpc, which makes the velocity unit $[v] = 977.8$ km/s. In these units, the disk parameters are: mass $M_d = 0.25$, total radius $R_d = 21$, scale length $R_e = 3.5$, scale height $z_e = 0.3$, and the value of $Q$ at $R = 8.5$ is 1.20. The halo parameters are: mass $M_h = 1.0$, total radius $R_h = 50$, core radius $R_c = 1.0$, and velocity dispersion $\sigma_h = 0.13$.

### 2.2. Infall Model

We allow the halo to virialize initially by holding the disk frozen for a few crossing times. The disk and halo are then allowed to evolve together for about 10-12 crossing times before the gas is brought into the picture. The gas is configured as a rectangular slab of uniform density and constant velocity dispersion. In accordance with the typical random velocities of molecular clouds, the velocity dispersion of the gas is assumed to be $\sim 10$ km/s. The density of the slab determines how much gas is added to the galaxy per unit time, and hence is kept low enough to ensure that the gas does not fall in faster than the disk can handle. The maximum infall rate which the disk can withstand without significant heating was experimentally determined to be a few percent (2-5 %) of its mass per crossing time. The length of the initial rectangle of gas is therefore proportional to the total mass of the gas. With the origin at the center of the galaxy and the galactic disk in the xy plane, the gas slab is placed slightly above the xy-plane and parallel to the $y$-axis. It is then given a velocity (i.e. each gas particle is given the same orbital velocity) in the negative $y$ direction which is an appropriate fraction of the Keplerian velocity $\sqrt{GM/R}$, where $M$ is the mass of the galaxy, and $R$ is the distance of the gas slab from the $y$-axis. This initial velocity (and hence the initial angular momentum) is set to be small enough that the gas falls into the galaxy's potential well, and large enough that it does so within a Hubble time. The direction of the velocity is also set to ensure that the gas orbit is retrograde with respect to the primary disk's rotation.

We follow the motion of the gas and disk particles until all the gas settles into the plane of the disk. Although it is possible to comment on the size, orientation and general appearance of the counterrotating disk formed, the simple sticky particle scheme cannot be used to study the more complex physics that must come into play after this point, such as shocks, star formation, heating by supernovae, etc. There is also the implicit assumption that no gas particles are lost to star formation prior to their forming a counterrotating disk. As a result, we can only study the gross properties of the infall process, such as the time-scale of counterrotating disk formation, the overall size and shape of the counterrotating disk formed, and the thickening of the primary disk.

### 2.3. Gas-Rich Dwarf Merger Model

Dwarf galaxies range from dwarf ellipticals with ellipsoidal mass distributions to dwarf irregulars with unknown mass distributions. There is no "garden-variety" dwarf galaxy model as such, and even properties such as mass, radius and gas content vary widely. Hence we decided to use a spherical Plummer model to represent our dwarf galaxy. In this model, the mass density function is assumed to be a polytrope of index 5:

$$\rho(r) = \left(\frac{3}{4\pi}\right) \frac{M}{R^3[1+(r/R)^2]^{5/2}}$$

where $M$ is the total mass, and $R$ determines the size (radius) of the sphere. Actually, the three mass components - gas, stars and dark matter - form concentric Plummer spheres with the gas sphere being the smallest and the dark matter being the largest. Initial positions and velocities of the equal-mass particles in each Plummer model are determined according to a scheme described in the literature (Aarseth *et al.* 1974).

The dwarf galaxy (secondary) is placed in the plane of the primary's disk at a large enough distance from the primary that its gravitational influence on the disk orbits is negligible; the initial separation is a little more than 3 halo radii (150 kpc). The secondary is also given a small non-zero velocity which places it in a retrograde orbit around the primary's disk so that if a second disk is formed it will rotate counter to the original disk.

We carried out three simulations of mergers between our spiral galaxy and a gas-rich dwarf galaxy. The primary difference among these is the mass of the dwarf galaxy. This is the quantity which has the most significant impact on the outcome of the simulation. The important parameter values are summarized in Table 1 for the three runs dubbed S1, S2 and S3.

### 2.4. Numerical Code

For the gravity solver, we have adapted a C-language version of the hierarchical TREE code (Barnes & Hut 1986) for the infall problem and implemented it on the Cray Y-MP at the Ohio Supercomputer Center. Fundamental changes to the TREE code were necessary to optimize and in particular vectorize it so that it would run efficiently on the Cray (Hernquist 1990, Makino 1990). Performance characteristics of the TREE code have been discussed by Hernquist (1987), and an error analysis (Barnes & Hut 1989) has shown that errors can be made quite small with judicious choice of parameters, especially for large particle numbers ($> 10^4$). To ensure good accuracy, we use quadrupole moments for all simulations, and an opening angle (clumping parameter) $\theta = 0.8$.

In all the simulations presented here, we use 32k particles to represent the disk, 16k particles for the halo, and up to 20k particles for the gas (depending on its mass). Each gas particle is about 1.3 times as massive as a disk particle. The disk crossing time is $\sim 100$ Myr, so the disk is held frozen for $t = 0 - 500$ Myr to allow the halo to virialize. Following this, the disk and halo are allowed to evolve together for $t = 500 - 1500$ Myr. The gas is then introduced at $t = 1500$ Myr.

One concern in simulations of disks is the minimization of two-body encounters so as to keep the disk as relaxation-free as possible for the duration of the simulations. Although relaxation rates in tree codes are comparable to those in PM codes with similar force resolution (Hernquist & Barnes 1990), and the number of disk particles we chose is enough to keep the disk relatively relaxation-free for the longest of our simulations in the absence of any other effects (Hockney & Eastwood 1989), the discreteness of the halo heats up the disk significantly as time progresses. Since our halo is 4 times as massive as the disk, each halo particle is 8 times as massive as a disk particle. Hence halo particles passing through the disk can disrupt the disk and alter its vertical structure much like the $\sim 10^6 M_\odot$ black holes postulated by Lacey & Ostriker (1985). A test simulation showed that doubling the number of halo particles (to 32k) reduces the increase in disk thickness by 40-50% on average. The increased heating of the disk with radius for our more discrete halo is consistent with the predicted swing amplification of the halo discreteness noise by the disk (Sellwood 1989).

The effect of the discrete halo on the disk also necessitates the choice of a rather high softening length, $\epsilon = 1.0$ kpc. This limits our vertical resolution to 1 kpc, and changes in thickness which amount to less than this cannot be considered significant. This limited resolution is still enough to enable us to identify broad trends in the behavior of the disk, however.

We have chosen a time-step of 5 Myr. Since the resolution is set by $\epsilon$, the time-step is small enough that the average particle (whose velocity $\sim 0.2$) does not move more than the spatial resolution length in one time-step. Furthermore, since the minimum collision radius for sticky-particle collisions is $r_{\min} = \epsilon/2$ (see below), this also ensures that very few if any collisions are missed between successive time-steps, since the relative velocity between colliding particles is usually much less than 0.2.

### 2.5. Sticky Particle Method

The sticky particle method we adopted includes features from several previous implementations (Brahic 1977, Schwarz 1981, Negroponte & White 1983). We use the standard scheme for binary inelastic collisions, which conserves linear and angular momentum and dissipates kinetic energy by reducing and reversing the relative radial velocity of the colliding particles. The tangential component of the relative velocity is left intact. The final velocities of colliding particles $p$ and $q$ are given by (Brahic 1977):

$$\mathbf{v}'_p = \frac{1}{2}\left\{\mathbf{v}_p + \mathbf{v}_q + [\mathbf{v}_{pq} \cdot \mathbf{i}]\alpha_c \mathbf{i} + (\mathbf{v}_{pq} \cdot \mathbf{j})\mathbf{j}\right\},$$
$$\mathbf{v}'_q = \frac{1}{2}\left\{\mathbf{v}_p + \mathbf{v}_q - [\mathbf{v}_{pq} \cdot \mathbf{i}]\alpha_c \mathbf{i} - (\mathbf{v}_{pq} \cdot \mathbf{j})\mathbf{j}\right\}$$

where $\mathbf{v}_p$ and $\mathbf{v}_q$ are the initial velocities, $\mathbf{v}_{pq} = \mathbf{v}_p - \mathbf{v}_q$, $\mathbf{i}$ is a unit vector in the radial direction from $p$ to $q$, $\mathbf{j}$ is a unit vector along the tangential component of $\mathbf{v}_p - \mathbf{v}_q$, and $\alpha_c$ is the coefficient of restitution ($-1 < \alpha_c < 0$ for an inelastic collision). The energy lost after one such collision is $(1 - \alpha_c^2)[\mathbf{i} \cdot (\mathbf{v}_p - \mathbf{v}_q)]^2$. The stickiness of the gas particles is adjusted by varying the coefficient of restitution $\alpha_c$; the number of collisions can also be controlled by the minimum and maximum collision radii.

We allowed each particle to be involved in at most one collision per time-step in order to make the collisions at any given time-step independent of each other. The second particle chosen for the collision is the one nearest to the current particle which meets the other criteria necessary for a collision to occur and which has not already been paired with another particle. Hence multiple collisions are not possible and only binary collisions are handled. This was necessary for the collisions to be computed in parallel. For this reason our gas is probably not excessively dissipative even for rather low values of $\alpha_c$ ($\sim -0.25$). We chose $\alpha_c = -0.5$ for the simulations reported here.

A collision between two particles occurs when the distance between them is less than a pre-defined "collision radius" ($r_c$), which varies with the density to make the number of collisions roughly independent of density. This avoids the overlapping of clouds in dense regions (Negroponte & White 1983). To avoid collisions in very dense regions (where $r_c$ is small and can cause a runaway increase in density), we define a minimum collision radius $r_{\min}$ whose value is set to half the gravitational softening length (or resolution length) $\epsilon$; to limit the search volume around each particle in which to look for potential collisions, a maximum collision radius $r_{\max}$ was used (we adopted a value of 3.0 for $r_{\max}$). It was also necessary to have a minimum relative K.E. cutoff $E_{\min}$ so that particles bound to each other do not engage in too many successive low-energy collisions. This limit was specified by setting the minimum relative collision velocity to be 0.010. If the relative K.E. between two colliding particles is less than $E_{\min}$, the particles undergo an elastic collision instead of an inelastic one. Thus particles do not interpenetrate.

Finally, to the standard collision model we added the refinement necessary to distinguish between particles moving toward each other and those moving away from each other. Thus collisions are avoided if the particles are separating (as implied by the angle between the relative radial velocity and the relative radial position vector), even if they are within the collision radius.

## 3. Results and Discussion

The list of input parameters and initial conditions that can have a tangible effect on the outcome of the simulations is quite long. For dissipational infall, the parameters which we decided to test include the infall rate of the gas, the mass of the galactic halo, the "stickiness" of the gas (as specified by the coefficient of restitution $\alpha_c$), and the magnitude and direction of the initial angular momentum of the gas (the former being specified by the initial velocity of the gas, the latter by the inclination of the infall trajectory to the plane of the disk). For the gas-rich dwarf merger, the primary input parameter tested is the mass of the dwarf galaxy.

For all simulations, results are shown starting at $t = 1.5$ Gyr, which is the point at which the gas/dwarf-galaxy is introduced.

### 3.1. Episodic Infall

Our *episodic* infall is really *periodic* infall, since the episodes are equally spaced in time. The formation of a counterrotating gas disk is illustrated in Figs. 1a and 1b. We introduce a new slab of gas after every 1.5 Gyr. The slab is placed slightly (5 kpc) out of the plane of the primary disk. The mass of each slab is 8% of the mass of the disk. The face-on (top) view (Fig. 1a) shows the gas forming a disk considerably smaller than the primary stellar disk, and the edge-on view (Fig. 1b) indicates that the thickness of the gas disk is comparable to that of the stellar disk.

The thickness of the primary disk is studied more quantitatively by plotting the mean half-thickness as a function of radius at different epochs during the simulation (Fig. 2). This is done by calculating the mean $z$-coordinate at each $(x, y)$ location by binning the $z$ values in bins of size 1 kpc$^2$. The average of the mean $z$-coordinates is then computed for each value of $R$. This ensures that any tilt or warp in the disk plane is accounted for, although there is some error introduced due to the fact that the thickness is measured vertically (parallel to $z$-axis) even though the disk may be tilted. Since we are more concerned with the trend in the thickness rather than its precise value, we have neglected this error.

There is a significant increase in the thickness of the disk, especially in the outer regions, as the simulation progresses. By $t = 9.0$, or 7.5 Gyr after the infall begins, the disk thickness has increased by a factor of 5-6 at some radii. The mass of the gas accumulated in the counterrotating disk by this time is about 40% of the primary disk's mass. As more gas is added, the thickness increases gradually and the heating of the disk progresses inwards towards the center, as more and more gas finds its way there.

In the meantime, the primary disk also develops a bar which gets more pronounced as more gas falls in, since the colder gas is more susceptible to bar formation. The final velocity fields of the primary disk and the gas (Fig. 3) confirm that the gas is counterrotating and also indicate that it has lost most of its velocity dispersion due to collisions and hence is significantly colder than the primary disk.

Once there is some counterrotating gas in the disk, successive episodes of infall can build up the counterrotating disk faster because the gas gets rapidly captured by the disk. The incoming gas particles collide with the ones already in the disk and hence get decelerated enough to be instantly inducted into the counterrotating flow. The heightened collision rate inhibits the "flinging" normally caused by tidal stripping, thereby facilitating the quick capture of the gas.

### 3.2. Continuous Infall

The discrete infall events in the periodic scheme have a somewhat *ad-hoc* nature to them. The gas is more or less "spoon-fed" to the spiral galaxy in a "palatable" form every so often. Although we do not know enough about the nature of real gas infall, a more realistic situation perhaps is one in which all of the gas is initially placed in a configuration which will allow it to make its own way to the galactic disk over an extended period of time (to keep the infall rate sufficiently low), the only stipulation being that it should do so on a retrograde orbit. To simulate such *continuous* gas infall, we make the initial rectangular slab of gas sufficiently long that its infall time is close to a Hubble time ($\sim 12$ Gyr). The mass of the slab is set to 80% of the mass of the primary disk, so that the counterrotating disk formed will be almost as massive as the primary disk.

In our first test of continuous gas infall, we made the cross-section of the rectangular slab the same as it was for periodic infall. The results in this case are shown in Fig. 4. It is clear that most of the length of the slab breaks up into individual clumps well before it falls into the galaxy. The clumps formed are large enough to have an appreciable effect on the disk, with the largest of them approaching the sizes of dwarf galaxies, with masses of the order of $10^9$ M$_\odot$. Even after tidal stripping, the cores of these are dense enough to remain intact and spiral into the center of the galaxy due to dynamical friction, which is proportional to the mass of the clumps to first order

(Binney & Tremaine 1987, eq. 7-26):

$$t_{\rm fric} \sim \frac{r_{\rm i}^2 v_{\rm c}}{GM} \sim \frac{1}{M}$$

where $t_{\rm fric}$ is the time it takes for an object with mass $M$ and initial distance $r_{\rm i}$ to spiral in under dynamical friction to the center of an isothermal mass distribution whose circular velocity is $v_{\rm c}$. A simple calculation for $M = 10^9$ M$_\odot$ and a maximum initial impact parameter of 50 kpc yields $t_{\rm fric} \simeq 6$ Gyr for $r_{\rm i} = 20$ kpc (the edge of the primary disk, where $v_{\rm c} \simeq 180$ km/s).

After about 4.5 Gyr (by $t = 6.0$), we see a counterrotating disk starting to form. Most of the gas has fallen in by about $t = 10.0$, and the final size of the counterrotating disk formed is quite small ($\lesssim$ half the size of the primary disk), as would be expected with most of the gas particles being clumped together very closely in phase-space and subjected to dynamical friction by the time they come close to the primary disk. The effect on the primary disk of "digesting" these clumps of gas is measured in terms of the mean half-thickness as a function of radius for different epochs during the infall (Fig. 5).

Clearly, the clumping of the gas particles poses a serious problem for the formation of a counterrotating disk whose size is comparable to that of the primary disk, and it also has a detrimental effect on the tranquillity of the primary disk. To minimize the clumping, we tried substantially reducing the density of the slab by increasing its cross-section but keeping its length the same to preserve the infall rate. The increased dynamical time in this case makes it much more difficult for the gas particles to coalesce into clumps before they fall into the galaxy. The results are shown in Fig. 6. It is evident that the clumping is much less pronounced than in Fig. 4, and the counterrotating disk formed in this case is much larger than before. It is almost the same size as the primary disk. Thus if the gas initially occupies a larger volume in phase-space, it will tend to form a larger counterrotating disk.

Another factor which can affect the size of the counterrotating disk formed is the magnitude of the initial angular momentum of the gas. Fig. 7 shows the result of increasing the initial velocity of the gas in the positive $x$ direction by a small amount (20% of $y$-velocity) so that the initial pericenter distance (and hence the angular momentum) is larger than before. The effect on the size of the counterrotating disk is remarkable, and we now get a counterrotating disk which is actually larger than the primary disk. Thus small changes in the initial angular momentum of the gas can have a significant impact on the scale length of the resulting counterrotating disk. Although both the initial phase-space density and the angular momentum of the gas affect the size of the counterrotating disk, the former is more important because it keeps the gas clump-free and therefore less susceptible to dynamical drag and also less likely to disrupt the primary disk.

Small differences in the inclination of the infall trajectory (in other words the initial direction of the angular momentum vector), appear to be less important to the final outcome, although highly inclined trajectories have been shown to have significant dynamical consequences (Sofue 1994). We do not see an appreciable difference in the outcome or the time-scale when the initial $z$-position of the gas is altered by 5 kpc in either direction (corresponding to an inclination of 1 in 10). Also, as the edge-on views in Fig. 7 indicate, the plane of the galactic disk itself changes as the infall progresses. This means that the gas which initially precesses to the plane of the primary disk must again realign itself as that plane itself shifts. The inner gas is aligned with the primary disk plane since it has had more time to do so, whereas the outer gas is still tilted with respect to it. This differential precession of the counterrotating disk is also clearly apparent in the lower half of Fig. 7, and a classic integral-sign warp develops as a result.

Beyond $t = 12.0$ Gyr, the counterrotating disk becomes more and more centrally concentrated. In our simple model which does not include shocks, star formation or re-heating by supernovae, it is not too surprising that the cool gas eventually makes its way to the center as it loses energy in collisions. This "overcooling" is a problem that has been observed also in simulations of dissipative galaxy formation (Navarro & White 1994), in which the primary disk formed is more centrally concentrated than real spiral disks due to cool gas making its way to the center.

The counterrotating gas disk is also more susceptible than the primary disk to bending instabilities such as warps. This is due to the coldness of its velocity field, as shown in Fig. 8. The kinetic energy dissipated by the gas in collisions causes it to be significantly colder than the primary disk (stellar) particles.

In order to study the role played by the collisional energy dissipation, we ran a simulation with the collisions turned off. The behavior of this collisionless fluid was dramatically different; in the time it takes for a collisional fluid to form a counterrotating disk ($\sim$ 7-9 Gyr), the collisionless fluid particles had not even begun to form a disk-like configuration. Most of them were on wildly eccentric orbits which took them far beyond the neighborhood of the primary disk rather than spiraling in towards the center of the disk. Thus tidal stripping and dynamical friction by themselves are not enough to produce a counterrotating disk within a Hubble time, and a mechanism for redistributing angular momentum, provided by inelastic collisions in this case, is necessary.

The initial angular momentum of the gas is the primary determinant of the time required to form the counterrotating disk. Besides tidal torques, net gas angular momentum can only be lost to the disk and halo particles through dynamical friction, and hence is a liability when the gas tries to form a counterrotating disk. When the initial $y$-velocity imparted to the gas is doubled, with everything else staying the same, the time required for the gas to fall in and circle the disk is at least doubled. Other factors, such as the stickiness of the gas or the inclination of the infall trajectory, are less important in determining the time-scale. A less sticky gas does not necessarily take longer to form a counterrotating disk. If $\alpha_c = -0.5$ instead of $-0.25$, the gas is less clumpy and more dispersed, but it takes about the same time to form a counterrotating disk, although the disk has a larger scale length and scale height. The initial height of the gas has a slightly more pronounced effect on the time-scale of the counterrotating disk formation, since the gas must precess to the plane of the disk first. Collisional dissipation facilitates this task, however, and it does not take very long for the gas to reorient its angular momentum vector.

A more massive (and hence more extended) halo is also instrumental in setting the time-scale of formation for the counterrotating disk. It provides a deeper potential well for the gas to fall into, so the gas can have a higher-energy orbit and the infall can proceed faster without making the angular momentum too high to form a counterrotating disk. Our initial attempts at continuous infall were unsuccessful because the halo was only twice as massive as the disk ($M_h = 0.5$), and the maximum relative velocity of the gas was too low to allow the infall process to complete within a Hubble time.

One of the necessary constraints on a successful model for counterrotating disk formation is that the original prograde disk must remain relatively unperturbed by the addition of retrograde gas whose mass is comparable to its own mass. This condition can only be met if the gas is added a little at a time. Simulations of dissipationless mergers have shown that a disk can be substantially heated by the acquisition of a satellite whose mass is only a few percent of the disk mass (Quinn 1987). In the case of dissipational infall, more mass can be added without disrupting the existing disk, although there is still a limit to how much gas mass can be added per dynamical time. We find that the heating of the disk increases sharply when the gas mass exceeds a few percent of the disk mass per crossing time; this is largely because the gas tends to form large clumps regardless of the value of $\alpha_c$. The large clumps disrupt the disk as discussed above. The disk also forms a bar in response to the infalling gas, and the bar gets more prominent as the infall rate increases. The sensitivity of the disk to the amount of gas falling in constrains the infall rate and prevents the rapid formation of a counterrotating disk if the primary disk is to remain undisturbed.

### 3.3. Dwarf Merger

Although some of the clumps of gas in the continuous infall case above had masses comparable to dwarf galaxies, an actual merger with a gas-rich dwarf is still worth investigating separately, because a real dwarf galaxy contains stars and dark matter in addition to gas. Dwarf galaxies also come in a range of sizes, and we wanted to find out if it was possible to obtain a counterrotating disk without severely disrupting the primary disk for any range of dwarf masses.

It is quite impossible to distinguish between the different mass constituents of both galaxies in a black and white image, so it is necessary to display the important components separately to study their behavior. The gas and stars in the dwarf galaxy, and the disk stars in the spiral galaxy, are the three constituents of interest. The evolution of the halos of both galaxies is not directly relevant to the formation of a counterrotating disk and hence is not shown.

The evolution of the spiral disk and the gas in the dwarf galaxy for run S1 is quite rapid. The stronger mutual gravitational attraction, combined with stronger tidal effects and increased dynamical friction on the massive dwarf, causes the two galaxies to merge within a relatively short time (Fig. 9). The dwarf is massive enough to survive the primary's tidal field, and sinks to the center of the primary relatively unscathed. As a result, there is no chance for a counterrotating disk to form in this nearly head-on merger.

In run S2, the mass of the dwarf is reduced by a factor of 3 by halving the dark halo mass as well as the luminous mass. The dwarf gas evolution along with the primary disk is shown in Fig. 10 (top view). A counterrotating disk is formed in about 5-6 Gyr.

Finally, in run S3, the mass of the dwarf is halved again. Fig. 11 shows the evolution of the dwarf gas, and Fig. 12 shows the stars in the dwarf. The formation of a tidal tail each time the secondary makes a close pass past the primary shows how mass is gradually stripped from the dwarf and forms a counterrotating disk in the spiral. The gas forms a more flattened disk than the stars (not shown). Although a counterrotating disk is eventually formed in this case also, it takes significantly longer to form as the less massive dwarf is subject to gravitational "flinging" (due to tidal effects) several times before it totally merges with the primary. The total time taken to form the counterrotating disk is about 9 Gyr, which is prohibitively long if several such mergers are required to form a massive counterrotating disk. The effect of the influx of gas and stars on the primary's disk is studied by comparing the thickness of the disk at various epochs as the simulation progresses (Fig. 13). From the disk thickness plots shown in Fig. 13, it is clear that even with such a small dwarf mass (S3), the disk still suffers considerable damage in the outer portions, although the thickening is less in the inner half of the disk for S3.

The final velocity fields of the primary disk, the gas counterrotating disk, and the stellar counterrotating disk are compared for runs S2 and S3 in Fig. 14. The more massive dwarf results in a smaller, more centrally concentrated counterrotating disk, but the velocity field of the stars is dominated by random motions. In general, the stars do not acquire an ordered retrograde motion because of their collisionless evolution. This is seen more clearly in the angular momentum plots (Fig. 15), which show the mean angular momentum of the dwarf's stars to be much lower than that of the gas. The mass profiles plotted in Fig. 16 confirm that the stars in the dwarf do not have a disk-like (exponential) mass profile.

To summarize the situation with dwarf mergers, if the dwarf is too massive ($M_{\rm Dw} \gtrsim 0.2 M_{\rm Sp}$), it merges rapidly with the primary and no counterrotating disk is formed. If the dwarf is too small ($M_{\rm Dw} \lesssim 0.05 M_{\rm Sp}$), the counterrotating disk takes $\gtrsim 9$ Gyr to form, and since several such mergers will be required to produce a substantial counterrotating disk, there is not enough time to form one. For an intermediate dwarf mass, a small counterrotating disk forms within 5-6 Gyrs and two or three such events could conceivably produce a massive counterrotating disk within a Hubble time. The biggest problem with such a merger, however, is the thickening of the disk due to the effect of the dark matter and stars in the dwarf galaxy, most of which do not accompany the gas in forming a counterrotating disk. The behavior of the gas is consistent with simulations of gaseous mergers and spiral accretion (Sofue 1994) in which the stripping and accretion of gas proceeds at a much greater pace than for the stars, especially for retrograde mergers.

As noted in §2.4, the thickness of the primary disk increases with time, albeit much more slowly, even when the disk and halo are allowed to evolve together *in isolation* (in the absence of any gas infall or merger). The disk also shows traces of a bar and a tilt in this case. The thickening of the disk due to gas infall or a dwarf merger must therefore be compared with the thickening of an isolated disk, *not* with the initial pre-interaction disk. Fig. 17 compares the thickness of the disk at $t = 1.5$ (initial) with its thickness at $t = 9.0$ for an isolated galaxy as well as for dissipational infall and a dwarf merger. Whereas the thickness increases only slightly (and only in the outer half of the disk) for dissipational infall, the entire disk shows a dramatic increase in thickness (an order of magnitude) for the dwarf merger. In fact, with our limited resolution, it is not possible to say whether dissipational infall does cause a non-zero increase in the disk thickness.

### 3.4. Comparison with Previous Work on Mergers

Results of equal mass mergers cannot generally be extended to unequal mass mergers (Villumsen 1982); hence we restrict ourselves to mergers between spirals and dwarfs. There has been very little work to date on modelling such mergers, and practically none on retrograde mergers. Most studies of "gas-rich mergers" concentrate on mergers between a gas-rich disk galaxy with a gasless dwarf (e.g. Mihos & Hernquist 1994, Hernquist & Mihos 1995). Simulations of dissipationless mergers between spiral galaxies and smaller satellites (Quinn 1987, Quinn *et al.* 1993) have shown the dramatic impact that even a very small satellite, whose mass is only a few percent of the primary's mass, can have on the disk of the primary. The disk thickness increases by a factor of $\sim$ 2-3 for a satellite mass of 4% of the disk mass. The disk also spreads radially as a result of the spiral response to the perturbation due to the satellite. Our simulations provide confirmation of the radial and vertical spreading of the disk even in the case of a retrograde merger. As Figure 18 shows, there are fewer particles near the center and more particles at larger $r$ and $z$ for the post-merger disk.

To get a better idea of the role played by the gas, we made all the particles dissipationless in run S3 and compared the results with those obtained for S3. We found no appreciable difference in the thickness of the disk between the dissipationless and dissipational mergers. Of course, no counterrotating disk is formed in the totally dissipationless merger, since the collisionless gas particles now behave the same as the stars in S3. In the final analysis, although the presence of gas affects the behavior of the dissipationless particles to some extent (the stars tend to clump where the gas does), it does very little to alleviate the disruption of the disk caused by the stars and dark matter. The situation might be different if there was a lot more gas in the dwarf galaxy and little or no dark matter. This is consistent with an analytical study of spiral-dwarf mergers by Toth & Ostriker (1992) which concludes that the gas in the dwarf will have a negligible impact on the heating of the disk unless the dwarf is almost entirely ($> 90\%$) gaseous.

Another observation from simulations of dissipationless mergers is that denser, massive cores in the secondary show up as dense cores in the primary after the merger (Balcells & Quinn 1990). This is also demonstrated in Figure 14, which shows that the secondary's stars form a denser core in S2 (intermediate mass dwarf) than in S3 (low-mass dwarf).

## 4. Conclusions

Our experiments with dissipational infall indicate that the accretion of a counterrotating gas disk whose mass is comparable to that of the existing stellar disk is possible without excessive damage to the existing disk if the gas is acquired gradually in small batches. The severe restrictions on the mass that can be accreted in the form of a satellite galaxy through dissipationless mergers (Quinn 1987, Quinn *et al.* 1993) are much less stringent in the case of gas infall, and the ability of the gas to dissipate energy and redistribute angular momentum allows it to form a counterrotating disk within a Hubble time. The gas infall can be in the form of successive episodes of infall, or a pseudo-continuous stream (since the gas breaks up into clumps anyway). The self-gravity of the gas, which has been taken into consideration in our simulations, is an important factor in determining both the clumpiness of the gas and its effect on the existing disk. A clumpy gas will tend to heat up the primary disk and form a small, centrally concentrated counterrotating disk. The initial angular momentum of the gas crucially affects both the size of the disk formed and the time required for its formation.

Collisional dissipation enables the gas to quickly redistribute and reorient its angular momentum, allowing it to settle into the plane of the disk to form a counterrotating disk within a Hubble time. A collisionless gas is unable to do this, and hence dissipationless infall may be ruled out as a probable mechanism for forming counterrotating disks.

A merger with a gas-rich dwarf galaxy can produce a small counterrotating disk whose mass is a small fraction of the primary disk mass, but it would be very difficult to accumulate an extensive counterrotating disk whose mass is comparable to that of the primary disk with this process. Although we tried only three representative cases of a merger with a gas-rich dwarf galaxy, it is clear that obtaining a massive counterrotating disk by this method is a difficult proposition. The need to form such a disk within a Hubble time with minimal disruption of the primary disk is virtually impossible to satisfy with one or more mergers of this kind. The disruptive influence of the dwarf's dark halo and to some extent the stellar matter in the dwarf is strong enough to heat the primary disk considerably, causing a sharp increase in its thickness. Extremely gas-rich dwarf galaxies which have little or no dark matter could succeed in producing counterrotating disks, but unless these represent the majority of dwarf galaxies, the probability of forming massive counterrotating disks is still significantly diminished.

The bleak outlook for dwarf mergers means that a potentially prolific mechanism for producing counterrotating disks is lost. Since dwarf galaxies may form naturally in tidal tails of encounters between giant galaxies (Hernquist & Barnes 1992), dwarf mergers would have increased significantly the probability of obtaining counterrotating disks in spirals, especially in clusters of galaxies. Without them, we are left with dissipational gas infall as the primary producer of massive counterrotating disks in spirals. The conditions for such infall to succeed, namely a relatively clump-free gas maintaining a retrograde orbit and a low infall rate, must persist over nearly a Hubble time, and this leads us to believe that they must not be a frequent phenomenon even though most spiral galaxies accrete gas throughout their lives. Although we would expect to see a few more because our vision has improved considerably, it is not likely that many more will be seen.

The sticky particle scheme only allows us to qualitatively address basic questions, such as whether gas infall or a gas-rich dwarf merger are likely routes to counterrotating disks. It is not competent enough to permit inquiry into the detailed characteristics of the gas disk (scale length, height) or the intricacies of its effect on the primary disk. We expect to carry out such a detailed study after incorporating SPH gas dynamics in our code.

A.R.T. thanks Jens Villumsen for guidance during the early stages of this project, and acknowledges useful comments from Lars Hernquist, Joshua Barnes, Isaac Shlosman, David Merritt and Hans-Walter Rix. Support for this work was provided by a NYI award to B.S.R. (NSF grant AST-9357396), NASA grant NAG 5-2864, and an Ohio Supercomputer Center Research Grant (PAS825) for computer time on the Cray Y-MP. We wish to thank the O.S.C. staff, particularly David Ennis and Tim Rozmajzl, for all their help. Finally, we thank Richard Pogge, David Weinberg and the anonymous referee for valuable comments and suggestions on the manuscript.

| Run No. | Dwarf Mass | Dwarf GMF[a] | Dwarf MLR[b] | Dwf/Sp TMR[c] | Dwf/Sp LDR[d] |
|---|---|---|---|---|---|
| S1 | 0.30 | 0.6 | 3.0 | 24% | 40% |
| S2 | 0.10 | 0.6 | 2.0 | 8% | 20% |
| S3 | 0.05 | 0.6 | 2.0 | 4% | 10% |

Table 1: Parameter values for gas-rich dwarf merger simulations.

---

[a]Gas Mass Fraction - fraction of luminous mass in gas particles.

[b]Mass to Light Ratio - ratio of total mass to luminous (stars + gas) mass.

[c]Total Mass Ratio - ratio of total mass of dwarf to total mass of spiral.

[d]Luminous to Disk mass Ratio - ratio of luminous mass in dwarf to disk mass in spiral.

Fig. 1a.— Face-on view of counterrotating disk formation by periodic infall. The galactic (primary) disk is shown only in the first panel for clarity. The galactic halo is not shown. Each slab of gas is 8% as massive as the disk. The square panels are 200 kpc on each side. Note that all panels are not equally spaced in time. Time is in Gyr.

Fig. 1b.— Edge-on view of periodic gas infall. The galactic disk is shown only in the first panel. The gas is initially 5 kpc above the plane of the disk.

Fig. 2.— (*a*) Top (left panels) and side (right panels) views of the primary (galactic) disk for different epochs during periodic infall. The width of each square panel is 40 kpc. (*b*) The mean half-thickness of the disk as a function of radius for each epoch. Each panel shows the effect of adding more gas in steps of 8% of the primary disk's mass, so that by $t = 9.0$, the total mass of gas in the counterrotating disk is about 40% of the primary disk mass.

Fig. 3.— Velocity fields of the disk and counterrotating gas at $t = 8.5$. Only a small subset ($\sim 6\%$) of the total number of particles is shown for clarity. The top panel shows the primary disk, rotating counterclockwise, and the bottom panel shows the gas, rotating clockwise.

Fig. 4.— (*a*) Top and (*b*) side views of counterrotating disk formation by continuous gas infall. The initial configuration of the gas is a long, thin rectangular slab of uniform density.

Fig. 5.— The thickness of the disk measured for different epochs during continuous infall.

Fig. 6.— (*a*) Top and (*b*) side views of continuous infall with a thicker (less dense) initial slab of gas than in Fig. 4. The counterrotating disk formed is slightly larger than the primary disk.

Fig. 7.— (*a*) Top and (*b*) side views of continuous infall with a thick initial slab of gas and a higher initial gas velocity than in Fig. 6. This gives the gas more angular momentum and the result is a larger counterrotating disk. For $t > 9.0$ Gyr, the image scale is doubled so that each panel width is 100 kpc (instead of 200 kpc).

Fig. 8.— Velocity fields of the disk and counterrotating gas at $t = 9.0$ Gyr for the continuous infall simulation in Fig. 7. Only a small subset ($\sim 12\%$) of the total number of particles is shown for clarity.

Fig. 9.— (*a*) Top and (*b*) side views of a merger with a massive gas-rich dwarf galaxy. The two galaxies merge rapidly, before there is a chance to form a counterrotating disk. Both the stellar and gas particles in the dwarf are shown. The primary disk is shown only in the first panel.

Fig. 10.— Formation of a counterrotating disk by a merger with an intermediate mass gas-rich dwarf galaxy (run S2). Only the gas particles in the dwarf are shown. The luminous mass (stars + gas) in the dwarf is one-fifth the mass of the primary disk. The counterrotating disk formed is less than half the size of the primary disk. The primary disk is shown only in the first two panels.

Fig. 11.— Merger with a gas-rich dwarf whose luminous mass is only about one-tenth of the mass of the primary disk (run S3). Only the gas particles are shown and the primary disk is shown only in the first two panels. The counterrotating disk formed in this case is almost as extended as the primary disk, but much less massive. It also takes much longer to form (compared to S2).

Fig. 12.— The evolution of the stars in the dwarf galaxy with time for run S3. By the end of the simulation, the stars are still dispersed over a large region and have quite eccentric orbits (see velocity fields in Fig. 14).

Fig. 13.— The thickness of the disk at different epochs for runs S2 and S3. In the outer parts of the disk, the heating of the disk is quite drastic, with an increase in thickness of more than 10 times.

Fig. 14.— The velocity fields of the primary disk, secondary gas and secondary stars (from left to right) are compared for S2 (top panels) and S3 (bottom panels).

Fig. 15.— The mean $z$-component of the angular momentum plotted as a function of radius for dwarf mergers S2 and S3. The angular momenta of both the gas (dotted line) and the stars (dashed line) belonging to the dwarf galaxy are directed opposite to that of the primary's disk (solid line).

Fig. 16.— The mass profile of the stars (dashed line) and gas (dotted line) compared to the mass profile of the primary disk (solid line) for runs S2 and S3.

Fig. 17.— A comparison of the initial thickness (solid line) of the primary disk with its thickness at $t = 9.0$ Gyr for four different simulations: an isolated galaxy (no gas infall or merger), periodic gas infall, continuous gas infall, and a gas-rich dwarf merger (S3).

Fig. 18.— The final ($t = 9.0$ Gyr) vertical density profile of the primary disk shown at various radii for the isolated disk, following periodic and continuous infall, and after dwarf merger S3. The density is averaged over azimuth for a given radius.